\documentclass[aps,prd,notitlepage,nofootinbib,superscriptaddress,twocolumn,floatfix]{revtex4-2}

\pdfoutput=1

\usepackage{graphicx}
\usepackage{soul}
\usepackage{subfigure}
\usepackage{latexsym}
\usepackage{amssymb}
\usepackage{amsmath}
\usepackage{color}
\usepackage{array}
\usepackage{mathrsfs}
\usepackage{xparse}
\usepackage{bbding}
\usepackage{enumitem}
\usepackage{pifont}
\usepackage{enumerate} 
\usepackage{color}
\usepackage{mathrsfs}
\usepackage{xcolor}
\usepackage{changes}
\usepackage[colorlinks=true,linkcolor=blue,urlcolor=blue,citecolor=blue]{hyperref}

\begin{document}

%%%%%%%%%%%%%%%%%%%%%%%%%%%%%%%%%%%%%%%%%%%%%%%%%%%%%%%%%%%%%%%
 \newcommand{\bq}{\begin{equation}}
 \newcommand{\eq}{\end{equation}}
 \newcommand{\bqn}{\begin{eqnarray}}
 \newcommand{\eqn}{\end{eqnarray}}
 \newcommand{\nb}{\nonumber}
 \newcommand{\lb}{\label}
 
\newcommand{\La}{\Lambda}
\newcommand{\va}{\scriptscriptstyle}
\newcommand{\be}{\nopagebreak[3]\begin{equation}}
\newcommand{\ee}{\end{equation}}

\newcommand{\ba}{\nopagebreak[3]\begin{eqnarray}}
\newcommand{\ea}{\end{eqnarray}}

\newcommand{\la}{\label}
\newcommand{\n}{\nonumber}
\newcommand{\su}{\mathfrak{su}}
\newcommand{\SU}{\mathrm{SU}}
\newcommand{\U}{\mathrm{U}}

\def\be{\nopagebreak[3]\begin{equation}}
\def\ee{\end{equation}}
\def\ba{\nopagebreak[3]\begin{eqnarray}}
\def\ea{\end{eqnarray}}
\newcommand{\f}{\frac}
\def\rmd{\rm d}
\def\lp{\ell_{\rm Pl}}
\def\d{{\rm d}}
\def\fe{\mathring{e}^{\,i}_a}
\def\fw{\mathring{\omega}^{\,a}_i}
\def\fq{\mathring{q}_{ab}}
\def\t{\tilde}

\title {Quantum Damping of Cosmological Shear: A New Prediction from Loop Quantum Cosmologies}  

\author{Wen-Cong Gan}
\email{ganwencong@jxnu.edu.cn}
\affiliation{School of Physics, Jiangxi Normal University, Nanchang, 330022, China}

\author{Leila L. Graef}
\email{leilagraef@id.uff.br} 
\affiliation{Instituto de F\'isica, Universidade Federal Fluminense, 24210-346 Niteroi, RJ, Brazil}

\author{Rudnei O. Ramos}
\email{rudnei@uerj.br}
\affiliation{Departamento de F\'isica Te\'orica, Universidade do Estado do Rio de Janeiro, 20550-013 Rio de Janeiro, RJ, Brazil}

\author{Gustavo S. Vicente}
\email{gustavo@fat.uerj.br}
\affiliation{Faculdade de Tecnologia, Universidade do Estado do Rio de Janeiro, 27537-000 Resende, RJ, Brazil}

\author{Anzhong Wang \thanks{Corresponding author}}
\email{Anzhong$\_$Wang@baylor.edu; the corresponding author}
\affiliation{GCAP-CASPER,  Department of Physics $\&$ Astronomy, Baylor University, Waco, Texas 76798-7316, USA}

%%%%%%%%%%%%%%%%%%%%%%%%%%%%%%%%%%%%%%
\begin{abstract}

We study the dynamics of the Bianchi~I universe in modified loop 
quantum cosmology (mLQC-I) and uncover a robust mechanism for isotropization: 
the shear is dynamically suppressed after the bounce and decays 
rapidly in the quantum post-bounce regime, independently of the 
equation of state of standard matter sources. This naturally drives 
the Universe toward a homogeneous and isotropic expanding phase 
without fine-tuning. Our results show that mLQC-I provides a  
new quantum-gravitational mechanism for suppressing anisotropies, 
absent in other bounce models.

\end{abstract}
%%%%%%%%%%%%%%%%%%%%%%%%%%%%%%%%%%%%%%%%

\maketitle
 
%%%%%%%%%%%%%%%%%%%%%%%%%%%%%%%%%%%%%%%%
{\em Introduction}---Loop quantum gravity (LQG) provides a leading 
nonperturbative and background-independent approach to quantum 
gravity~\cite{Ashtekar:2004eh,Thiemann:2007pyv,Rovelli:2014ssa}. Its 
symmetry-reduced formulation, loop quantum cosmology 
(LQC)~\cite{Ashtekar:2011ni,Agullo:2023rqq,Li:2023dwy}, has enabled 
concrete and potentially testable cosmological predictions. A hallmark 
result is the replacement of the big-bang singularity by a nonsingular 
quantum bounce at Planckian densities, while general relativity (GR) is 
recovered at low curvatures. Different quantization prescriptions within 
LQG, however, can lead to distinct effective dynamics, allowing qualitatively 
new features in the bounce and post-bounce evolution. 

Standard
LQC was originally formulated under two key assumptions: The \emph{minisuperspace} 
approximation, whereby homogeneity and isotropy are imposed prior to quantization 
and the use of a classical identity relating the Euclidean and Lorentzian 
terms~\cite{Ashtekar:2011ni,Agullo:2023rqq,Li:2023dwy}, allowing the  
Hamiltonian to be expressed solely in terms of the Euclidean sector. 
In full LQG, however, the Euclidean and Lorentzian parts are quantized differently, 
and symmetry reduction and quantization do not generally commute~\cite{Ashtekar:2004eh,Thiemann:2007pyv,Rovelli:2014ssa}.
To address  these issues, an alternative quantization procedure 
more faithful to LQG was proposed in~\cite{Yang:2009fp}, in which the two 
sectors are quantized separately, leading to a distinct effective theory 
known as modified loop quantum cosmology (mLQC-I)~\cite{Li:2021mop}. Unlike 
other modified LQC models, mLQC-I  admits both bottom-up~\cite{Yang:2009fp,Assanioussi:2018hee} 
and top-down derivations directly from LQG~\cite{Dapor:2017rwv,Han:2020uhb,Han:2022rsx}, 
making it a  particularly compelling realization within the LQG framework.
While the qualitative predictions of LQC, such as the replacement of the 
big bang by a quantum bounce, remain robust, mLQC-I exhibits markedly 
different dynamics in the contracting phase, including an asymmetric 
evolution around the bounce~\cite{Li:2018fco,Li:2018opr,Li:2019ipm,Saini:2018tto,Saini:2019tem,Saeed:2024xhk}. 
This motivates extending the analysis to anisotropic spacetimes such as 
Bianchi~I, in order to test the robustness of mLQC-I against shear 
and to explore its implications for isotropization.
Bianchi spacetimes classify all \emph{simply transitive} 
homogeneous three-geometries into nine types,
with FLRW models as special highly symmetric cases, making them a 
natural starting point for early-universe cosmology. Preliminary 
mLQC-I studies along these lines already exist, including the effective 
Hamiltonian for Bianchi~I spacetime \cite{Garcia-Quismondo:2019kav,Garcia-Quismondo:2019dwa}.

In the broader landscape of cosmological bounce models, including matter and 
ekpyrotic bounces \cite{Khoury:2001wf,Khoury:2003rt, Lehners:2008vx,Battefeld:2014uga,Brandenberger:2016vhg} 
and quantum bounces \cite{Ashtekar:2011ni,Agullo:2023rqq,Li:2023dwy}, a persistent 
challenge is the growth of anisotropies during the contracting phase. For generic 
initial conditions, anisotropy is dynamically expected;  the shear scales as 
$\Sigma^2/a^6$, where $\Sigma$ is a constant and $a$ the mean scale factor. 
This scaling causes anisotropies to grow faster than standard matter or radiation 
components, effectively acting as a stiff fluid. Unless suitably suppressed, 
they can  dominate the dynamics and leave large imprints on the post-bounce universe, 
in tension with observational constraints. Identifying mechanisms that control anisotropies 
across the bounce remains a central open problem. Classically, this difficulty 
is reminiscent of the Belinskii–Khalatnikov–Lifshitz %(BKL) 
conjecture \cite{Belinsky:1970ew}, 
arising from Bianchi VIII and IX models,
which asserts that near a spacelike singularity time derivatives dominate over 
spatial gradients, reducing the asymptotic dynamics to ordinary differential 
equations. It is therefore essential to explore how anisotropies evolve when the 
dynamics is governed not by the standard LQC, but by its modified version mLQC-I, 
where the quantization of the Hamiltonian more faithfully reflects the full LQG structure.

In this Letter, we investigate the impact of quantum geometric effects in mLQC-I 
on collapsing Bianchi I trajectories, with particular emphasis on singularity 
resolution and anisotropy dynamics. 
Our analysis is restricted to Bianchi I geometries; extensions to Bianchi VIII and IX, 
where chaotic mixmaster behavior competes with quantum damping, are left for 
future work. The Bianchi I spacetime provides an ideal testing ground: it is 
sufficiently simple for both analytical and numerical analysis, yet rich enough 
to capture anisotropic generalizations of the FLRW universe 
(see, e.g.,~\cite{Chiou:2006qq,Chiou:2007sp,Bojowald:2007ra,Ashtekar:2009vc,Corichi:2009pp,Cailleteau:2009fv,Garay:2010sk,Martin-Benito:2010dge,Gupt:2012vi,Singh:2011gp,Gupt:2013swa,Liu:2012xp,Yue:2013kd,Linsefors:2013bua,Linsefors:2014tna,Bodendorfer:2014vea,Singh:2016jsa,Wilson-Ewing:2017vju,Alesci:2019sni,Agullo:2020uii,Agullo:2020iqv,Agullo:2022klq,McNamara:2022dmf,Motaharfar:2023hil,Brown:2024xta}). 
We demonstrate that, for generic initial conditions with nonvanishing shear, 
the quantum bounce not only replaces the classical singularity but also 
dynamically drives the shear to zero deep in the post-bounce quantum regime. 
As a result, the universe naturally isotropizes and evolves into a homogeneous, 
isotropic, and macroscopic state.
Our result is conceptually related to Wald’s cosmic no-hair theorem~\cite{Wald:1983ky}, 
whereby a positive cosmological constant (CC) enforces isotropy. Here, however, 
the isotropization and subsequent de Sitter–like expansion after 
the bounce emerge purely from quantum geometric effects, rather than from a 
prescribed cosmological constant, making the suppression of anisotropies a genuinely quantum-gravitational effect.

%%%%%%%%%%%%%%%%%%%%%%%%%%%%%%%%%%%%%%%%
{\em Bianchi I universe in mLQC-I} --- The  Bianchi I universe is described by the metric
$ds^2   = -N^2(t)dt^2 + \sum_{i = 1}^{3}{a_i^{2}(t) \left(dx^i\right)^2}$, 
where $N(t)$ is the lapse and $a_i(t)$ are the scale factors along the three spatial directions. 
Adopting the Ashtekar variables \cite{Ashtekar:1986}, the momenta are defined as  
$p_i \equiv \text{sgn}(a_i)\left|a_ja_k\right|L_jL_k\; (i \not= j\not=k)$,
where $L_i$ denotes the length of the fiducial volume $\mathcal{V}$ in the $x^i$-direction.
As the final results do not depend on $L_i$, without loss of generality, we set $L_i = 1$.
The $p_i$ satisfy the canonical relations 
$\left\{c_i, p_j\right\} = 8\pi G \gamma \delta_{ij}$, where $c_i$ 
denotes the connection component conjugate to $p_i$ and $\gamma$ is the Barbero-Immirzi parameter
(in all of our numerical results, we choose $\gamma \approx 0.2375$, 
as suggested from black hole thermodynamics in LQG \cite{Meissner:2004ju}).
Since the spatial hypersurfaces are non-compact and all fields are spatially 
homogeneous, we construct a Hamiltonian within the fiducial cell $\mathcal{V}$ 
and restrict all integrations to 
it~\cite{Ashtekar:2004eh,Thiemann:2007pyv,Rovelli:2014ssa}.

Moreover,  we focus on the effective dynamics of the Bianchi I universe for 
sharply peaked trajectories such that $a_i \ge 0$ \cite{Ashtekar:2009vc}.
The effective gravitational Hamiltonian ${\cal{H}}_{\text{eff}}^{\text{mLQC-I}}$ 
for the Bianchi I universe in the framework of mLQC-I was derived  
in~\cite{Garcia-Quismondo:2019kav,Garcia-Quismondo:2019dwa} 
and is given by Eq.~(\ref{HmLQCI}) in {\bf End Matter}. 
In this formulation, the polymerization parameters, which encode the 
holonomy modifications of the connection, are specified  by~\cite{Chiou:2006qq,Ashtekar:2009vc}
 $\bar\mu_{i} = \sqrt{ p_{i} \Delta/({p_{j}p_{k}})},\;
 (i \not= j \not= k)$,
 where $\Delta \equiv 4\pi \sqrt{3} \gamma \ell_{\text{Pl}}^2$ denotes the non-zero minimal area gap 
 of LQG~\cite{Ashtekar:2004eh,Thiemann:2007pyv,Rovelli:2014ssa}.
 These expressions are well justified geometrically and  correspond to the only viable prescription for 
 Bianchi I universe~\cite{Ashtekar:2009vc,Motaharfar:2023hil}. 
 
%%%%%%%%%%%%%%%%%%%%%%%%%%%%%%%%%%%%%%%%
{\em Shear in loop quantum cosmologies} --- To understand the role of shear 
in the Bianchi universe, we consider time-like geodesics with a unit tangential vector
$u^{\mu}$. Then, we have \cite{Ryan:1975jw}  
$\nabla_{\nu}u_{\mu} = \left(g_{\mu\nu} + u_{\mu}u_{\nu}\right)\theta/3 + \omega_{\mu\nu} + \sigma_{\mu\nu}$,
where $\theta, \; \omega_{\mu\nu}$ and $\sigma_{\mu\nu}$ denote, respectively, 
the expansion scalar, vorticity and shear tensors of the time-like geodesics. 
{}For a homogeneous universe, we have  $\omega_{\mu\nu} = 0$, $\theta = H_1 + H_2 + H_3$,   
and the shear scalar is given by 
\bqn
\lb{Sigma2}
\sigma^2  \equiv \sigma_{\mu\nu}\sigma^{\mu\nu}  = \left(H_{12}^2 + H_{23}^2 + 
H_{31}^2\right)/3 \equiv  {6\Sigma^2}/{a^6},
\eqn
where $H_{ij} \equiv H_i - H_j$, $H_i \equiv \dot{a}_i/a_i$, and 
$a^3\equiv v =a_1a_2a_3 = (p_1p_2p_3)^{1/2}$. 
Before we discuss the behavior of shear in mLQC-I, it is instructive to first consider it in LQC~\cite{Ashtekar:2009vc}.  The dynamical equations for the $p_i$ is 
given by Eq.~(\ref{piLQC}) in {\bf End Matter}.
Note that in writing Eq.~(\ref{piLQC}), and also throughout this Letter, we have 
set the lapse function to $N = 1$. Hence, the timelike coordinate $t$  denotes 
the cosmic time.  In addition, in writing down the last step of Eq.~(\ref{piLQC}), 
we used the definition  $p_i = a_j a_k$. Hence, in the classical limit (\ref{eq2}), 
Eq.~(\ref{piLQC}) yields 
$H_j + H_k = (p_jc_j + p_kc_k)/(\gamma v)$. Substituting it into Eq.~(\ref{Sigma2}), 
we obtain the well-known result \cite{Ryan:1975jw,Chiou:2007sp} 
$\Sigma^2  =  \left(\alpha_{12}^2 + \alpha_{23}^2 + \alpha_{31}^2\right)/18$,
where $ \alpha_{ij} \left[\equiv \left(p_{i}c_i - p_jc_j\right)/\gamma\right]$ are 
integration constants. 
We impose the initial conditions in the remote contracting phase, $t_i \ll t_B$, 
where $t_B$ denotes the bounce time(s) and $t_i$ is the initial time (note that 
there could be multiple bounces and in this case, we take $t_B$ as the time of 
the first bounce). Hence,  LQC is well approximated by the classical theory. 
Thus, we have $\Sigma^2_{\text{LQC}}(t_i) \simeq \Sigma^2_{\text{CL}}(t_i)$. 
Then, Eq.~(\ref{Sigma2}) tells us that $\Sigma^2$ approaches $\Sigma^2_{\text{CL}}(t_i)$ 
asymptotically after the bounce, where the classical limit is achieved. 

Now, let us turn to mLQC-I. In this case, Eq.~(\ref{piLQC}) must be replaced by 
Eq.~(\ref{pimLQCI}) in {\bf End Matter}.
Then, in terms of ${\sigma}^{i}_{jk}$, which is defined by Eq.~(\ref{pimLQCI}), we find
\bqn
\lb{eq8}
3\sigma^2     = \left({\sigma}^{1}_{23} - {\sigma}^{2}_{31}\right)^2  
+ \left({\sigma}^{3}_{12} - {\sigma}^{2}_{31}\right)^2 
+ \left({\sigma}^{3}_{12} - {\sigma}^{1}_{23}\right)^2. ~~~
\eqn
Again, assuming that at the initial time we have 
$\sigma^2(t_i) \simeq \sigma^2_{\text{CL}}(t_i)$, from the above equation we can 
see that $\sigma(t \gg t_B) \rightarrow 0$, whenever the system approaches its classical 
limit after the bounce(s), so that the condition given by Eq.~(\ref{eq2}) is satisfied, 
for which we have ${\text{sn}}_i(t) \rightarrow 0 \; (i = 1, 2, 3)$. Hence, irrespective 
of its initial magnitude, the shear is dynamically filtered by the quantum bounce and 
vanishes after the bounce(s), whereby a homogeneous and isotropic universe results. 

To explicitly demonstrate the above claim, we consider
$a_i(t) = a(t) e^{\theta_i(t)}$, and assume small deviations from isotropy, $\left|\theta_i\right| \ll 1$.
 Then, in the post-bounce region ($\dot{v} > 0$), 
$\theta_i$ satisfies Eq.~(\ref{master}),
which has the general solution ${\cal F}(t)\simeq A+B a(t)^{-3(\omega+1)}$, 
where $A= 3/[\sqrt{\Delta}(1+\gamma^2)]$
and   $B$ is another constant given by Eq.~(\ref{A20}). In deriving 
this expression, we have also assumed a general barotropic fluid with energy density 
$\rho = \rho_0/v^{1+\omega}$. Therefore, for any fluid with equation of state 
$\omega> -1$, we have ${\cal F}(t) \to A > 0$. Then,  the anisotropies vanish 
exponentially, $\theta_i \propto \exp(-A t)$. Since the damping originates 
entirely in the gravitational sector, the result is insensitive to the 
microphysics of the matter sector as long as $\omega>-1$. Under these same conditions, we find that
the shear scalar Eq.~(\ref{eq8}) behaves like 
\begin{equation}
\sigma^2 \propto e^{-2A t}, \; t \gg t_B,
\label{sigma2decay}
\end{equation}
where $A^{-1} \approx 0.8\; t_P$ for $\gamma \approx 0.2375$ defines the characteristic time of the shear decay after the bounce. Clearly, this happens well within the Planck regime.
This damping of the anisotropies arises from the asymmetric quantization of Euclidean and Lorentzian sectors in mLQC-I, which introduces an effective dissipative channel for anisotropies absent in the standard LQC.

Next, we confirm these results numerically without imposing the conditions $\left|\theta_i(t)\right| \ll 1$, and  show that, 
in the framework of mLQC-I, such isotropizations happen well within the deep quantum regime and occur independently of the collapsing matter fields. 
They remain valid even for initial conditions (taken in the contracting phase) 
corresponding to large anisotropies.
Consequently, all three spatial directions expand rapidly to macroscopic scales, 
producing a homogeneous and isotropic universe directly from the quantum epoch. 
This feature is unique to mLQC-I and not shared by other bounce models~\cite{Lehners:2008vx,Battefeld:2014uga,Brandenberger:2016vhg,Ashtekar:2011ni,Agullo:2023rqq,Li:2023dwy}.

%%%%%%%%%%%%%%%%%%%%%%%%%%%%%%%%%%%%%%%%
{\em Numerical results} --- With the  effective Hamiltonian
and polymerization parameters, we numerically solve the  
Hamiltonian equations $\dot Q = \left\{Q, \mathcal{H}_T\right\}$,
where $Q \equiv \left(p_i, c_i, \phi, \pi_{\phi}\right)$, and 
$\mathcal{H}_T \equiv \mathcal{H}^{\text{mLQC-I}}_{\text{eff}} + {\cal{H}}_M$. 
In particular, the dynamical equations for   $p_i(t)$ and $c_i(t)$ are given 
respectively by Eqs.~(\ref{pimLQCI}) and (\ref{cimLQCI}) in {\bf End Matter}. 
Here ${\cal{H}}_M$ is the matter Hamiltonian given by ${\cal{H}}_M = v N \rho_M$, 
where $\rho_M$ is the energy density of the matter field, while  $\phi$  
collectively denotes the matter field, being $\pi_{\phi}$  its moment conjugate. 
{}For a scalar field with potential $V(\phi)$, we have $\rho_M =  \pi_{\phi}^2/(2v^2) + V(\phi)$ and $\left\{\phi, \pi_{\phi}\right\} = 1$.  

The dynamics of the effective Hamiltonian equations is uniquely determined by 
their initial conditions. To see how to choose these conditions,  let us first 
consider the case in which the matter field is given by a scalar field. 
The generalization to other matter fields is straightforward. The dynamical 
system consists of eight first-order differential equations for the eight 
physical variables $(p_i, c_i, \phi, \pi_{\phi})$. In order to uniquely 
determine the evolution of the universe, we need eight initial conditions 
for each of these eight variables. However, due to the Hamiltonian constraint 
$\mathcal{H}_T \approx 0$, we actually need to specify only seven of them, 
while the eighth will be obtained from the Hamiltonian constraint.  Without 
loss of generality,  we first choose the initial conditions for 
$c_I(t_i), p_j(t_i), \phi(t_i), \pi_{\phi}(t_i)\; (I = 1, 2)$  at the initial 
time $t = t_i$, and then solve the effective Hamiltonian constraint  
to find $c_3^{\text{eff}}(t_i)$. It is clear that the reduced phase space 
consists of  all possible real values of 
$c_I(t_i), p_j(t_i), \phi(t_i),  \pi_{\phi}(t_i)$. However, to compare the 
resultant solutions with the corresponding classical ones, we choose them as 
their corresponding GR values, such that we have 
$c_I(t_i) = c_I^{\text{GR}}(t_i)$,  $p_j(t_i) = p^{\text{GR}}_j(t_i)$, $\phi(t_i) = \phi^{\text{GR}}(t_i)$ and $\pi_{\phi}(t_i) = \pi_{\phi}^{\text{GR}}(t_i)$.  
To fix $t_i$, we require that in $t_i$ the condition 
$\left|c_3^{\text{eff}}(t_i) - c_3^{\text{GR}}(t_i)\right| \ll 1$.
The conditions in $t_i$ must be satisfied, which means that the initial 
conditions must closely follow their classical values. Once the initial 
conditions are chosen, the equations of motion will uniquely determine the 
eight physical variables at any given time $t$.

%%%%%%%%%%%%%%%%%%%%%%%%%%%%%%%%%%%%%%%%%%%%%%%%%%%%%%%%% 
\begin{figure*}[!htb]
\subfigure[]
{\includegraphics[width=0.37\textwidth]{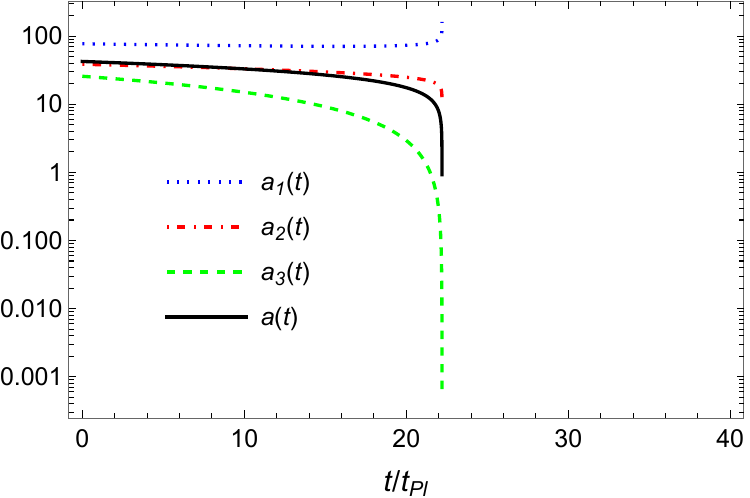}}
\subfigure[]
{\includegraphics[width=0.37\textwidth]{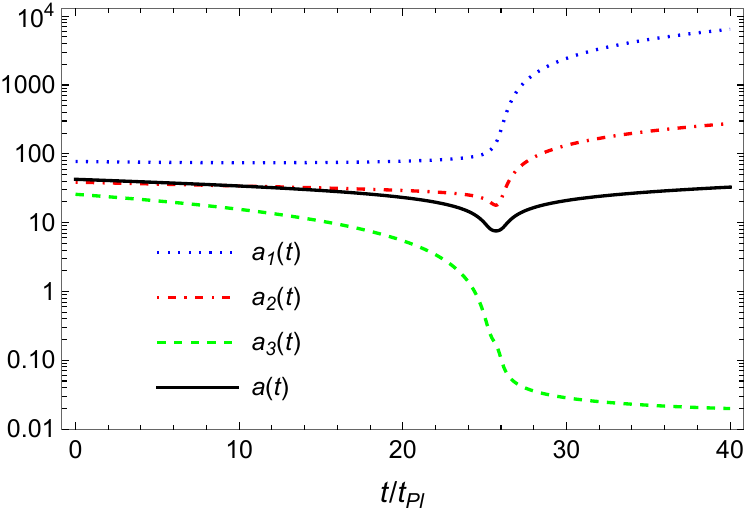}}
\subfigure[]
{\includegraphics[width=0.34\textwidth]{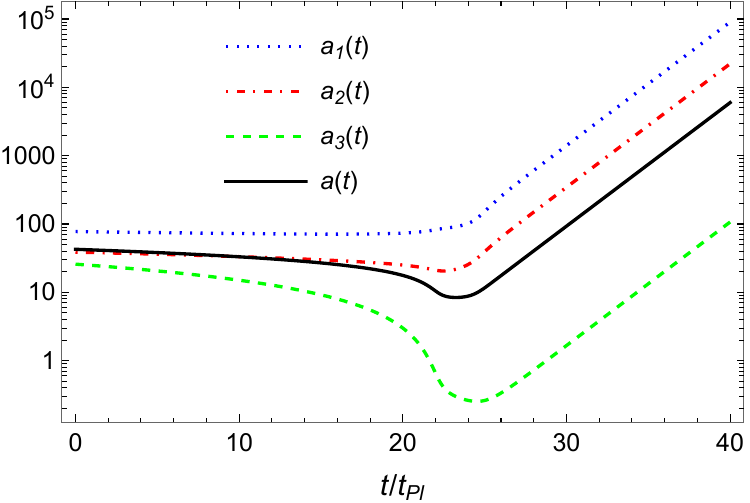}}
\subfigure[]
{ \includegraphics[width=0.39\textwidth]{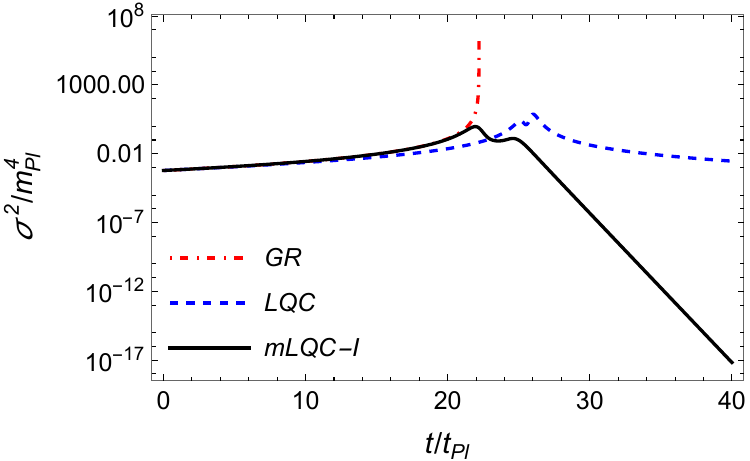}}
    \caption{Functions $a_i(t)$ and $a(t)$ for
    GR (panel a), LQC (panel b) and mLQC-I (panel c) for Bianchi I universe filled with dust.
    Shear scalar  
    is shown in panel d.
    Initial conditions are all imposed at $t = 0$.  }
    \label{fig1}
\end{figure*}
%%%%%%%%%%%%%%%%%%%%%%%%%%%%%%%%%%%%%%%%%%%%%%%%%%%%%%%%%%

With all of this in mind, we are ready to consider some representative examples. 
The first example corresponds to a spacetime filled with dust, for which we have 
$\rho_M =  {\rho_M^{(0)}}/{v}$, where ${\rho_M^{(0)}}$ is a constant. We choose the initial conditions as 
$\left(p^{\text{GR}}_1, p^{\text{GR}}_2, p^{\text{GR}}_3\right) = \left(10^3, 2\times 10^3, 3 \times 10^3\right)$,   
$\left(c^{\text{GR}}_1, c^{\text{GR}}_2\right) =\left(-0.13, -0.12\right)$, and $\rho_M =3.55 \times 10^{-5}$, where all quantities are expressed in units of Planck mass $m_{\rm Pl}$, while time is expressed in terms of Planck time $t_{\rm Pl}$. 
Then, using the Hamiltonian constraints, we find $c_3^{\text{GR}}=-0.243926$, $c^{\text{eff, LQC}}_3=-0.243947$, and  $c^{\text{eff, mLQC-I}}_3=-0.2439897$, 
from which we can see that the requirement $\left|c_3^{\text{eff}}(t_i) - c_3^{\text{GR}}(t_i)\right| \ll 1$ 
is well satisfied. 
Using the arbitrariness in setting the initial time, 
we also set $t_i = 0$.  The  dynamical equations of $p_i$ and $c_i$ are given, 
in {\bf End Matter},   by Eqs.~(\ref{piLQC}) and  (\ref{ciLQC})  for LQC, and 
by (\ref{pimLQCI}) and (\ref{cimLQCI})  for mLQC-I. The corresponding classical 
equations can be obtained from  Eqs.~(\ref{piLQC}) and (\ref{ciLQC}) taking the 
limits of Eq.~(\ref{eq2}).   In {}Fig.~\ref{fig1}, we show the results for the 
physical quantities ($a_i (t), a(t), \sigma^2(t)$) obtained respectively in GR, 
LQC and  mLQC-I. {}From {}Fig.~\ref{fig1} we can see that 
in all three cases the space is always collapsing initially, 
$\dot{a}_i(t) < 0\; (t_i < t < t_B)$, and a classical singularity develops at $t \simeq 22\, t_{\rm Pl}$ in GR. However, in both LQC and mLQC-I  it is replaced by a regular quantum bounce, 
with slightly different bouncing times: $t_{B}^{\text{LQC}} \simeq 26\, t_{\rm Pl}$ and
$t_{B}^{\text{mLQC-I}} \simeq 24\, t_{\rm Pl}$.  We also note that one 
of the $a_i$'s still remains of Planck size after the bounce  in 
LQC~\cite{Chiou:2006qq,Chiou:2007sp,Bojowald:2007ra,Ashtekar:2009vc,Corichi:2009pp,Cailleteau:2009fv,Garay:2010sk,Martin-Benito:2010dge,Gupt:2012vi,Singh:2011gp,Gupt:2013swa,Liu:2012xp,Gupt:2012vi,Yue:2013kd,Linsefors:2013bua,Linsefors:2014tna,Bodendorfer:2014vea,Singh:2016jsa,Wilson-Ewing:2017vju,Alesci:2019sni,Agullo:2020uii,Agullo:2020iqv,Agullo:2022klq,McNamara:2022dmf,Motaharfar:2023hil,Brown:2024xta}, while in  mLQC-I all three components $a_i$ grow rapidly after the bounce and 
soon reach their macroscopic sizes. 
In addition, in LQC the shear does not change significantly after the 
bounce~\cite{Ashtekar:2009vc,Chiou:2007sp}. However, in mLQC-I we see instead 
that the shear  decreases rapidly soon after the bounce, during which the system 
is still in the deep quantum regime. For the dust case, we find that it is decreasing exponentially as 
$\sigma^2(t) \simeq \sigma^2_0 \exp(- \Gamma t)$, 
where we find numerically that the decay rate is $\Gamma \simeq 2.498 \,m_{\rm Pl}$
for $t \gtrsim 26\; t_{\text{Pl}}$, as can be seen in {}Fig.~\ref{fig1} (d). 
This is a confirmation of the claim made above about the rapid vanishing of the shears after the bounce
and agrees quite well with the analytical estimate given by Eq.~(\ref{sigma2decay}), where $\Gamma=2A  \approx 2.49806 \, m_{\rm Pl}$.

To further demonstrate that the behavior given by Eq.~(\ref{sigma2decay}) is {\em universal} 
not only perturbatively but also non-perturbatively, we have studied  various  cases with different initial conditions and different matter fields
and found that the above results are robust. {}For example, in addition to the case of dust considered above, we have also considered the cases of a radiation field $\rho_R = {\rho_R^{(0)}}{v^{-4/3}}$, a free massless scalar field, as well as a scalar field with an ekpyrotic potential (EKP)~\cite{Cai:2012va,McNamara:2022dmf}, and a  polynomial chaotic potential (PCP)~\cite{Kallosh:2025ijd}, given by Eq.(\ref{eq13}), which involves two sets of parameters 
 $(V_0, p, \beta)$ and $(\alpha_{1}, \alpha_2, m)$. Choosing $\alpha_1 = 0.14$, $\alpha_2 = 6.644\times 10^{-3}$ and $m = 1.26 \times 10^{-6}\; m_{\rm Pl}$,  it was shown \cite{Kallosh:2025ijd} that
the PCP model fits recent cosmic microwave background observations very well~\cite{ACT:2025fju}. 
In the following, we shall choose $(V_0, p, \beta) = (0.00366, 0.1, 5)$ for the EKP model, 
to ensure $c^{\text{eff}}_3(t_i)<0$ (ensuring that our initial conditions set   
lead to collapse of the Bianchi I universe in the pre-bounce phase, 
$\dot{a}_i(t) < 0$ for $t_i < t < t_B$), while keeping the same choice for 
$\alpha_{1, 2}$ and $m$ as those adopted in \cite{Kallosh:2025ijd}. 
For such choices,  we find that in the case with an EKP potential, similar to the case shown 
in~\cite{McNamara:2022dmf}, two bounces appear, while in the other three cases only a single 
bounce exists. Like in the dust example, in all these four cases we still find that all $a_i(t)$ 
and $a(t)$ grow dramatically after the bounce,  $\dot{a}_i(t), \; \dot{a}(t) > 0$ 
for $t > t_B$,  and soon reach their macroscopic sizes.  
In {}Fig.~\ref{fig2},
we show $\sigma^2(t)$ for these four different cases, from which we can see that the 
exponential decrease all happens in the deep quantum regime, at which the conditions 
for the classical limit $\left|\mu_i c_i\right| \ll 1$ are not held yet. Note that the 
massless case is almost indistinguishable from the  polynomial chaotic case. Within our 
numerical errors,  we find that in all these cases $\sigma^2$ 
 follows the general behavior given by Eq.~(\ref{sigma2decay}) and with the same rate $\Gamma=2A$, 
confirming that this behavior
is {\em universal: shears approach zero exponentially with the same form
and exponent in the deep quantum regime. }

%%%%%%%%%%%%%%%%%%%%%%%%%%%%%%%%%%%%%%%%%%
\begin{figure}[h!]  
\includegraphics[width=0.37\textwidth]{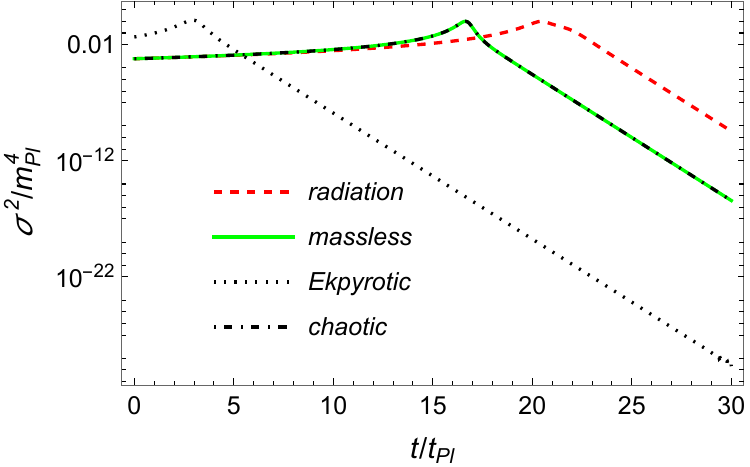}
\caption{Shear
in mLQC-I for 
a radiation fluid,  massless scalar field, or a scalar field with an ekpyrotic or a polynomial chaotic potential, 
with the same initial conditions.}
\lb{fig2}
\end{figure} 
%%%%%%%%%%%%%%%%%%%%%%%%%%%%%%%%%%%%%%%%%%

A natural question concerns how the quantum-generated de Sitter–like phase 
transitions to the classical FLRW regime \cite{Li:2018fco,Li:2018opr,Li:2019ipm}. 
Since this accelerated stage emerges 
dynamically from quantum geometry rather than from a fundamental cosmological constant, 
it need not persist indefinitely. A plausible mechanism is the dynamical relaxation of an effective 
vacuum energy through infrared backreaction of long-wavelength modes, as discussed in the 
context of the de Sitter space in ~\cite{Tsamis:1992sx,Tsamis:1994ca,Mukhanov:1996ak} and further analyzed 
in~\cite{Marozzi:2006ky,Abramo:2001dd,Losic:2005vg,Brandenberger:2018fdd,Woodard:2025smz,Glavan:2026pug}. 
In such scenarios, the cumulative effect of 
super-Hubble fluctuations shall gradually reduce the effective expansion rate, allowing radiation 
or matter to dominate and classical evolution to resume. 
Thus,  
our current results are consistent 
with a transient post-bounce accelerated phase dynamically connected to standard cosmology. 
{}Further discussion can be found in A.4 of {\bf End Matter}.

%%%%%%%%%%%%%%%%%%%%%%%%%%%%%%%%%%%%%%%%
{\em Conclusions}---
Using controlled analytical approximations supported by numerical integrations, 
we have identified a defining feature of the mLQC-I model: the exponential 
suppression of shear immediately after the quantum bounce. This universal 
isotropization mechanism operates independently of the content of collapsing matter
 and sharply distinguishes mLQC-I from other cosmological bounce scenarios~\cite{Ashtekar:2011ni,Lehners:2008vx,Battefeld:2014uga,Brandenberger:2016vhg,Ashtekar:2009vc,Motaharfar:2023hil}. 
Our results show that the emergence of a homogeneous and isotropic universe is a 
rapid and unavoidable outcome from the deep quantum regime, arising as an 
intrinsic property of quantum geometry in mLQC-I, rather than from additional assumptions or mechanisms.
This has important implications for both LQC and mLQC-I.
Although in both frameworks inflation occurs 
generically~\cite{Ashtekar:2011rm,Li:2019ipm}, large  shear  in the contracting 
phase could dominate the evolution of the Universe,   and hence inflation may not 
occur at all, after the shear effects are taken into account.
Indeed, recent work has shown that introducing an 
ekpyrotic phase to control anisotropies in the pre-bounce phase 
can severely restrict the region of initial conditions leading to successful 
inflation~\cite{Brown:2025hcb}. In contrast, in mLQC-I the shear problem is 
resolved dynamically by 
pure quantum geometric effects, without the need of provoking 
additional mechanism to suppress the shear. As a result, the naturalness of post-bounce inflation in mLQC-I persists, even if large shear exists  in the contracting phase. The dynamical relaxation of an effective 
vacuum energy through infrared backreaction of long-wavelength modes ~\cite{Tsamis:1992sx,Tsamis:1994ca,Mukhanov:1996ak} 
is a possibility to naturally lead to the end of inflation, whereby 
the standard cosmology is achieved, although it is important to note that the shear suppression demonstrated in this work occurs before and independently of the detailed mechanism responsible for the eventual relaxation of the accelerated phase. 

%%%%%%%%%%%%%%%%%%%%%%%%%%%%%%%%%%%%%%%%
\acknowledgements

W-C.G. would like to thank Drs. Cong Zhang, Hong-An Zeng, Chao Zhang, Bao-Fei Li and Shengzhi Li
for valuable discussions. He is supported by the National
Natural Science Foundation of China under the Grant No. 12405064, Jiangxi Provincial 
Natural Science Foundation under the Grant No. 20242BCE50055, and the Initial 
Research Foundation of Jiangxi Normal University.  L.L.G is supported by research 
grants from Conselho Nacional de Desenvolvimento Cientıfico e Tecnologico (CNPq),
Grant No. 307636/2023-2 and from the Fundacao Carlos Chagas Filho de Amparo a Pesquisa do Estado do Rio de
Janeiro (FAPERJ), Grant No. E-26/204.598/2024. 
R.O.R. is partially supported by research grants from CNPq, Grant No. 307286/2021-5, and FAPERJ, Grant
No. E-26/200.415/2026.
A.W. is partially supported by the US NSF  grant, PHY-2308845.

%%%%%%%%%%%%%%%%%%%%%%%%%%%%%%%%%%%%%%%%%%%%%  

%%%%%%%%%%%%%%%%%%%%%%%%%%%%%%%%%%%%%%%%%%%%%%

%\renewcommand{\theequation}{A\arabic{equation}}\setcounter{equation}{0}

\newpage
\appendix
\section{End Matter}

{\em A.1 Dynamical Hamiltonian Equations for GR, LQC and mLQC-I}---This appendix 
collects the effective Hamiltonians and equations of motion for GR, LQC, and mLQC-I, 
and provides the analytical derivation underlying the isotropization mechanism 
discussed in the main text. 
Throughout this Letter, Latin indices $i,j,k$ range from 1 to 3. We work in natural 
units with $\hbar=c=1$, where $\hbar$ and $c$ denote the reduced Planck constant and 
the speed of light, respectively. The Planck time and length are defined in terms of 
Newton’s constant $G$ as $t_{\rm Pl}=\ell_{\rm Pl}\equiv G^{1/2}$, and the Planck 
mass as $m_{\rm Pl}\equiv G^{-1/2}$.

The classical GR Hamiltonian is given by
\bqn
\lb{HclGRM}
{\cal{H}}_{\text{cl}}^{\text{GR}} &=&- \frac{N}{8\pi G \gamma^2 v}\left(c_1c_2p_1p_2 + c_1c_3p_1p_3
+ c_2c_3p_2p_3\right)\nb\\
&& + {\cal{H}}^{\text{M}}, 
\eqn 
where ${\cal{H}}^{\text{M}} \equiv v N \rho_M$ is the Hamiltonian for the matter 
content, with $\rho_M$ being the energy density of the matter field. {}From now 
on we will consider the lapse function $N=1$. The effective Hamiltonian of LQC 
can be obtained from the classical one by replacing $c_i$ in (\ref{HclGRM}) with  
\bqn
\lb{eqA3}
c_i \rightarrow \frac{\sin(\bar\mu_i c_i)}{\bar\mu_i}.
\eqn
 Then, the effective Hamiltonian equations in LQC can be obtained from the Hamiltonian equations,
$\dot{Q} = \left\{Q, \mathcal{H}^{\text{LQC}}_{\text{eff}}\right\}$.
In particular, we find
\begin{eqnarray}
%\bqn
\lb{piLQC}
\frac{\dot{p}_i}{p_i} &=& \frac{{\text{cs}}_i}{\gamma v} \left(p_j\; \mu\text{sn}_j  + p_k \; \mu\text{sn}_k\right)
= H_j + H_k, \\
\dot{c}_i &=& -\frac{1}{2\gamma p_i v} \left[
c_i p_i \, \mathrm{cs}_i \left(p_j \, \mu\mathrm{sn}_j + p_k \, \mu\mathrm{sn}_k\right) \right.
\nonumber \\
&& - 
c_j p_j \, \mathrm{cs}_j \left(p_i \, \mu\mathrm{sn}_i  + p_k \, \mu \mathrm{sn}_k\right)  \nonumber\\
&& - c_k p_k \, \mathrm{cs}_k \left(p_i \, \mu\mathrm{sn}_i  
+ p_j \, \mu\mathrm{sn}_j\right) 
\nonumber\\
&&+  p_i \, \mu\mathrm{sn}_i \; p_j \,\mu \mathrm{sn}_j
+ p_i \, \mu\mathrm{sn}_i\; p_k \, \mu\mathrm{sn}_k\nb\\
&&\left.
+ p_j \, \mu\mathrm{sn}_j\; p_k \, \mu\mathrm{sn}_k\right]
\nonumber \\
&&+  \frac{4 \pi G \gamma v}{ p_i} \left( \rho_M + 2 p_i \frac{\partial \rho_M}{\partial p_i} \right),
\label{ciLQC}
\end{eqnarray}
where $i \not= j \not= k$, ${\text{cs}}_i \equiv \cos(\bar\mu_i c_i)$, ${\text{sn}}_i \equiv \sin(\bar\mu_i c_i)$, and $\mu\text{sn}_i \equiv {\text{sn}}_i/{\bar\mu_i}$.
The classical GR equations are recovered from 
Eqs.~\eqref{piLQC}–\eqref{ciLQC} by taking the limit 
\bqn
 \lb{eq2}
 \lim_{\bar\mu_i \rightarrow 0}{\mu\text{sn}_i} 
 = c_i.
 \eqn
 
The effective Hamiltonian of mLQC-I is explicitly given 
by~\cite{Garcia-Quismondo:2019kav,Garcia-Quismondo:2019dwa}
\begin{eqnarray}
    \label{HmLQCI}
\mathcal{H}^{\text{mLQC-I}}_{\text{eff}} &=& \frac{N}{16\pi G v} \sum_{i,\, j \ne i}  p_i \; \mu\mathrm{sn}_i  \;  p_j\; \mu\mathrm{sn}_j\nb\\
&& \times
\left( 1 - \delta_\gamma \sum_{k \ne i, \, l \ne j} \mathrm{cs}_k  \mathrm{cs}_l \right)
 + {\cal{H}}^{\text{M}},
\end{eqnarray}
where we have defined $\delta_{\gamma} \equiv (1+\gamma^2)/(4\gamma^2)$. 
This effective Hamiltonian follows from a separate quantization of the Euclidean 
and Lorentzian terms, in contrast to LQC, and encodes genuine quantum-geometric corrections.
The  dynamical equation for $p_i$ in mLQC-I is  given by 
\begin{eqnarray}
\lb{pimLQCI}
\frac{\dot{p}_i}{p_i} = - \frac{\gamma}{\sqrt{\Delta}}
\left({\text{sn}}_i 
f^{\text{A}}_{ijk} +
{\text{sn}}_j 
f^\text{B}_{i,ik,jk} 
+
{\text{sn}}_k 
f^\text{B}_{i,ij,jk}
\right)
\equiv \sigma^{i}_{jk}, ~~~~~~~
\eqn
where 
$f^\text{A}_{ijk} 
\equiv 
\delta_{\gamma}[{\text{sn}}_i({\text{sn}}_j + {\text{sn}}_k)({\text{cs}}_j + {\text{cs}}_k)  + {\text{sn}}_j{\text{sn}}_k(2{\text{cs}}_i + {\text{cs}}_j + {\text{cs}}_k)]$
and 
$f^\text{B}_{i,jk,lm} 
\equiv 
{\text{cs}}_i[1 - \delta_{\gamma}({\text{cs}}_j + {\text{cs}}_k)({\text{cs}}_l + {\text{cs}}_m)]$. 
Similarly, we find that the equation of motion for $c_i$ is given by
\begin{eqnarray}
\lb{cimLQCI}
\dot{c}_i  &=& \frac{\gamma}{2p_i v} \big[
{c_i p_i} \,  
\left(p_j \,\mu\mathrm{sn}_j  f^\text{B}_{i,ik,jk} 
+ p_k \, \mu\mathrm{sn}_k  f^\text{B}_{i,ij,jk}
\right) \nonumber\\
&& - c_j p_j \,  
\left(p_i \, \mu\mathrm{sn}_i  
f^\text{B}_{j,ik,jk}
+  p_k \, \mu\mathrm{sn}_k  
f^\text{B}_{j,ij,ik}
\right) \nonumber\\
&& -
c_k p_k \,  
\left(p_i \, \mu\mathrm{sn}_i 
f^\text{B}_{k,ij,jk}
+  p_j \, \mu\mathrm{sn}_j  
f^\text{B}_{k,ij,ik}
\right) \nonumber\\
&& + p_j \, \mu\mathrm{sn}_j  p_k \, \mu\mathrm{sn}_k  
g^{-}_{ij,ik}
+
 p_i \, \mu\mathrm{sn}_i  p_k \, \mu\mathrm{sn}_k  
g^{+}_{ij,jk} \nonumber\\
&& +
 p_i \, \mu\mathrm{sn}_i  p_j \, \mu\mathrm{sn}_j  
g^{+}_{ik,jk}
\big] \nb\\
&& + 
\frac{4\pi G \gamma v}{p_i} \left( \rho_M + 2p_i \frac{\partial \rho_M}{\partial p_i} \right),
\end{eqnarray}
where 
$g^{\pm}_{ij,lk}
\equiv 
1 
- 
\delta_\gamma [(\mathrm{cs}_i + \mathrm{cs}_j)(\mathrm{cs}_l + \mathrm{cs}_k)
\pm
(\mathrm{cs}_i + \mathrm{cs}_j)
(
\bar{\mu}_l c_l \, \mathrm{sn}_l
\pm 
\bar{\mu}_k c_k \, \mathrm{sn}_k 
)
-(\mathrm{cs}_l + \mathrm{cs}_k)
(
\bar{\mu}_i c_i \, \mathrm{sn}_i
- 
\bar{\mu}_j c_j \, \mathrm{sn}_j 
)
].
$
In the main content, when we consider a scalar field $\phi$ with a potential $V(\phi)$ as the collapsing matter, we consider the cases  in which the potential takes the form
\bqn
\lb{eq13}
V(\phi) = \begin{cases}
 - \frac{2V_0}{e^{-\sqrt{\frac{16\pi G}{p}}\; \phi} + e^{\beta\sqrt{\frac{16\pi G}{p}}\; \phi}},  & \text{EKP}, \cr
    \frac{1}{2}m^2\phi^2\left(1 - \alpha_1\phi + \alpha_2\phi^2\right)^2, & \text{PCP}, \cr
\end{cases}
\eqn
where $(V_0, p, \beta)$ and $(\alpha_{1}, \alpha_2, m)$ are free parameters of the models.

{\em A.2 Anisotropies in mLQC-I}---Let us now consider the above equations for mLQC-I 
and expand them around the isotropic case. The matter content is taken to be a barotropic fluid with energy density $\rho_M=\rho_0/v^{1+\omega}$ and parameterize the line element for Bianchi I in the form 
\begin{equation}
    ds^2=-dt^2+a^2(t) \sum_{i=1}^{3} e^{2\theta_i(t)} (dx^i)^2,
\end{equation}
where $\sum_{i} \theta_i=0$. 
The variables $\theta_i$ parameterize deviations from isotropy, with $\theta_i=0$ corresponding exactly to the FLRW geometry.
Assuming perturbations around the FLRW universe, i.e.,
in the approximation of
small anisotropies, $\left|\theta_i\right| \ll 1$, 
to the leading order we find 
$a_i = a e^{\theta_i} = a (1 + \theta_i)$, 
$p_i = a^2 [1 +  (\theta_j + \theta_k)]$,
and $c_i =c(1+\delta c_i)$,
where $\sum_i \delta c_i=0$.
Defining $\bar{\mu}=\sqrt{\Delta}/a(t)$ and $b(t)=c(t)/a(t)$, then expanding the equations of motion to their first order, we obtain
\begin{eqnarray}
    \dot{v} &=&  
    \frac{3v \sin\bigl(2{\cal{B}}\bigr)}{2\gamma\sqrt{\Delta}} 
    \left[ (\gamma^2 + 1)\cos\!\bigl(2{\cal{B}}\bigr) - \gamma^2 \right], ~~~~~~~
    \label{dotv}
        \\
    \dot{b} &=&     \frac{3\sin^2\bigl({\cal{B}}\bigr)}{2\gamma\,\Delta} 
    \left[ \gamma^2\sin^2\!\bigl({\cal{B}}\bigr) - \cos^2\!\bigl({\cal{B}}\bigr) \right] - 4\pi G \gamma P, ~~~~~~\label{db}\\
\dot{\theta}_i &=& f(t) \left(\delta c_i - \theta_i \right),  \;\; \dot{\delta c}_i = g(t) \left( \delta c_i - \theta_i\right),
\label{anisoeqs}
\end{eqnarray}
with ${\cal{B}} \equiv \sqrt{\Delta}\; b$, $P=-\partial \mathcal{H}_M/\partial v=\omega \rho_M$, $\rho_M=\rho_0/v^{1+\omega}$, and $f(t)$ and $g(t)$   are defined, respectively, by
\begin{eqnarray}
\lb{eqA14}
 f(t)&\equiv&\frac{b}{32 \gamma}\left[4 (7 \gamma ^2+3) \cos (2{\cal{B}})\right.\nb\\
 && \left. -7 (\gamma ^2+1) \cos (4 {\cal{B}})-3(7 \gamma ^2-9)\right],\\
 \lb{eqA15}
   g(t) &\equiv&  f(t) +
    \frac{3  \sin ^2({\cal{B}})}{4\gamma  b \Delta }  \left[(\gamma ^2+1) \cos (2{\cal{B}})  -\gamma ^2+1\right]\nb\\
    && 
    -  \frac{3   \sin (2{\cal{B}})}{2 \sqrt{\Delta} \gamma}
    \left[(\gamma ^2+1) \cos (2{\cal{B}})-\gamma ^2\right] \nb\\
    && + \frac{4\pi G \gamma  \text{$\rho_{\rm 0}$} \omega } {b \,v^{\omega +1}}.
\end{eqnarray}
These equations show that the anisotropies are governed by the difference $(\delta c_i-\theta_i)$, whose evolution is entirely determined by the quantum-modified background through the functions $f(t)$ and $g(t)$.
Next, let us consider far from the bounce, $t \gg t_B$, but still in the quantum regime such that the classical pressure term in Eq.~\eqref{db} can be neglected. 
Then, Eq.~\eqref{db} has the nontrivial asymptotic fixed points given by
\bqn
\lb{asym}
b_{\text{asym}} = \left(0,  \pm \text{arctan}(1/\gamma)/\sqrt{\Delta}\right).
\eqn
It can be shown that when $b=0$ we have ($\dot{b}, \dot{v}) = (0, 0)$. Thus, it 
represents an attractive critic point of the phase space \cite{Li:2019ipm}.  
Therefore, in the rest of this Appendix, we shall not consider it further and 
focus ourselves on the cases  $b=\pm \arctan(1/\gamma)/\sqrt{\Delta}$, which give
\begin{eqnarray}\lb{dbv}
(\dot b,\dot v)=\left(0,\mp\frac{ 3 v}{(\gamma ^2  +1)\sqrt{\Delta}} \right). 
\end{eqnarray}
{}From the above expressions we can see that $b=-(1/\sqrt{\Delta}) \arctan(1/\gamma)$ is an attractor along the $b$-axis, and corresponds to an expanding solution. Around it,
$b$ approaches a constant and $v$ grows exponentially, such that 
\bqn
\lb{BB}
a(t) \simeq a_{0} e^{t/\delta}, \quad \delta \equiv {(\gamma ^2  +1)\sqrt{\Delta}},
\eqn
with  $a_0$ being a constant and corresponding to an accelerated, de Sitter–like 
attractor generated by the pure quantum geometric effects in the post-bounce regime.

{\em A.3 Master equations}---Combining the first-order anisotropy equations in 
Eq.~\eqref{anisoeqs}, we obtain a closed second-order evolution equation for $\theta_i$
\begin{align}\label{master}
    \ddot{\theta_i}+\mathcal{F}(t)\dot{\theta_i}=0,
\end{align}
where $\mathcal{F}(t)\equiv f-g-\dot{f}/f$, where $f$ and $g$ are given by Eqs.(\ref{eqA14}) and (\ref{eqA15}), respectively.
Working again far from the bounce, but still in the quantum regime, we find that the function $\mathcal{F}(t)$ can be well approximated by
\begin{align}
    \mathcal{F}(t) \simeq A+B\, a(t)^{-3 (\omega +1)},
\end{align}
with
\begin{equation}
\lb{A20}
A =\frac{3}{\delta},\;\;\;\;
B=
\frac{4\pi G\sqrt{\Delta}\rho_0}{4\gamma^2+1}
\left(1+19\gamma^2-20\gamma^4+20\gamma^2\omega\right).
\end{equation}
Thus, for $\omega > -1$, we have $\mathcal{F}(t) \approx A =
3/[(\gamma ^2  +1) \sqrt{\Delta}]$. Hence, Eq.~\eqref{master} has the general  solution 
\begin{align}
\theta_i \simeq C_i e^{-\mathcal{F}(t) t}+D_i,
\label{deltatheta}
\end{align}
where $C_i$ and $D_i$ are integration constants. The constants $D_i$ can be gauged away by rescaling $x^i \rightarrow e^{D_i} x^i$, and  do not contribute to shear, which depends only on $\dot{\theta}_i$. {}From Eq.~(\ref{deltatheta}) we can see that the anisotropy $\theta_i$ decreases exponentially
and becomes very small as $t \gg A^{-1} \simeq 0.8 t_{\text{Pl}}$, that is,  far from the bounce, but still in the quantum regime.

{\em A.4 Illustrative mechanism for relaxation of the 
post-bounce accelerated phase}---The quasi–de Sitter phase emerging in 
mLQC-I is generated dynamically by quantum geometry rather than by a 
fundamental cosmological constant. It is therefore natural to expect that 
this accelerated stage may be transient once additional dynamical effects 
are taken into account. In de Sitter space, it has been 
argued that infrared backreaction of super-Hubble fluctuations can effectively 
reduce the expansion rate~\cite{Tsamis:1992sx,Tsamis:1994ca,Mukhanov:1996ak}. 
As modes exit the Hubble horizon during exponential expansion, 
long-wavelength contributions accumulate and may modify the effective stress–energy tensor, 
leading to a gradual relaxation of the expansion. Such effects can become especially 
important  in the Planck regime~\cite{Marozzi:2006ky}. 
As an illustration, for a scalar field with quartic self-interaction 
$V(\phi)=\lambda \phi^4/4!$, perturbative analyzes indicate that the effective 
Hubble parameter can receive secular corrections of the schematic form~\cite{Abramo:2001dd}
\begin{equation}
H_{\rm eff} = H \left[ 1 - {\cal D}(Ht)^n + \cdots \right],
\end{equation}
where ${\cal{D}} = \lambda^2 G\Lambda/(2^73^4\pi^5)$, $n=4$ and $H$ is the Hubble parameter without the back reaction $({\cal{D}} = 0)$. In general, they are model-dependent and  
determine the 
strength of back reaction.   Such corrections suggest that the accelerated phase gets slowed down over time, 
allowing the Universe to approach a radiation- or  matter-dominated regime.
Recently, it was also shown that even when non-linear perturbations are taken into 
account, similar conclusions are obtained~\cite{Brandenberger:2018fdd,Woodard:2025smz,Glavan:2026pug}, which further suggest that the back reaction of the infrared perturbations is relevant, especially in the Planck regime, where
the Hubble horizon $L_H (\equiv (aH)^{-1} = H^{-1}e^{-Ht}$) decreases not only exponentially but also with the Planck rate
$H = {\cal{O}}(t_{\text{Pl}}^{-1})$, and physical modes  rapidly exit the Horizon. Then,  their effects quickly accumulate and  become non-negligible, so that any consistent treatment of the problem should take such effects into account. 

\end{document}